# Investigations of Smart Health Reliability

Sharlet Claros, Wei Wang, Ting Zhu

University of Maryland, Baltimore County
sclaros1@umbc.edu    ax29092@umbc.edu    zt@umbc.edu

*Abstract*—A balanced investigation into the reliability of wireless smart health devices when it comes to the collection of biometric data under varying network/environmental conditions. Followed by a program implementation to begin introductory analysis on measurement accuracy and data collection to gauge the reliability of smart health devices.

## I. Introduction / Motivation

Patient care and evaluation has revolutionized in this decade thanks to advancements in smart health devices and wireless communications We now have a wide range of access to wireless devices that are able to collect critical health data such as temperature, weight, blood pressure, and more. The field of smart health and biotechnology have become of increased interest more recently especially due to the COVID-19 pandemic and the development of AI/ML applications. More specifically, the expansion of telemedicine and telehealth patient monitoring has made people begin to switch over to wireless biometric collecting devices a priority for continuous care more than ever. However, it has long been a question if this is a safe/viable avenue of healthcare. The collection of biometric data through wireless or "smart health" devices facilitates these applications in telemedicine and patient monitoring as well as opens avenues for increased innovation in AI/ML datasets and applications in biotechnology. Therefore, it is significant to be able to determine with certainty whether any particular smart health device can deliver consistent and accurate data under varying conditions and network strengths. One of the biggest issues regarding smart health most recently has been surrounding the accuracy of the data collected, especially under different environmental and network conditions. In this paper I had originally planned to investigate and test the reliability of some of the most common metrics taken in by smart health devices through a simulated collected sample of biometric data. Then, apply the collected data to a programmed patient portal to be able to conduct introductory comparative analysis to be able to determine whether a particular smart health device is delivering consistent data. However, due to unforeseen health circumstances I instead utilize simulated data and keep my original outlined plan for the experimental process for reference.

## II. Related Work

Upon review of related work and approaches that have been accomplished in the field of wireless, biometric and AI technologies [10-35], I came across the following three investigations into smart health reliability.

### A. Wearables Reliability and Mortality Risk

The reliabilities and risks of smart devices are evaluated in multiple papers [1, 4-6]. For example, as referenced [1], this source references a study conducted on fitness trackers and smart watches that indicated that they were more effective predictors of mortality than patient surveys or other methodologies. Despite being a study surrounding a sample of some smart health devices, it differs from this approach in this paper in that this paper originally planned to investigate the reliability of smart health devices while the source describes a study conducted to determine if fitness trackers and smartwatches are viable predictors for patient mortality.

### B. Smart Device Accuracy for Measuring Physical Activity

Multiple researchers are also focusing on the accuracy of measuring physical activities [2, 7-9]. As referenced [2], this method explores the accuracy of smart devices to measure physical activity in daily life. This approach is much closer to the purpose and scope of what this paper originally planned to evaluate, which is smart health reliability. The key difference is the main focus of the evaluation, this source evaluates devices that are strictly used for physical activity, while this paper originally planned to evaluate accuracy for any sample of smart health devices and analyze accuracy across the board in smart health.

*C. Smart Watch Use for Health and Wellness*

As referenced [3], this study takes an in depth focus on smart watches and every functionality related to smart watch devices for health and wellness. The study certainly has a similar focus in that health functionality is a central focus, but the paper has a different scope than the investigation that this paper planned to take on. This paper originally planned to test reliability across all smart devices for accuracy under varying environments rather than just smart watches.

III. DESIGN

In response to the concerns surrounding the reliability of smart health devices, my original plan was to lay out an experiment in which I evaluate various smart health devices under simulated data collection and mathematically/ programmatically determine whether each device is reliable based on the experiment results.

This approach had to be modified due to unforeseen health circumstances that did not allow for physical data collection and needed to be switched to simulated collections instead. But I have decided to keep the original concept of this investigation in this paper to be able to express what the original intentions and utility is along with a logical conclusion for future use and applications.

The devices in this particular sample that were going be evaluated included a smart blood pressure monitor, a smart thermometer, and a smart watch. The biometric data that was to be collected with each device was temperature, blood pressure and heart rate specifically. The smart watch is capable of additional measures to track health, but for simplicity, the collected data was limited to only heart rate for the smart watch. For each of the measurements that were to be taken with the smart health devices, there was also a plan to take hand/traditionally taken measurement for each of the evaluated metrics that provided a baseline measurement to gauge smart health device accuracy.

To be able to properly examine the performance of a sample of smart health devices, evaluation was planned to be done under various environmental and network conditions to test reliability in performance. More specifically, data collection was to be partitioned into indoor, outdoor, 5GHz, and 2.4GHz collections. Data was planned to be collected over a span of two weeks for a single participant, with two data collections a day under a balance of indoor/outdoor and 5GHz/2.4GHz collections. For each smart health device data collection, there was also to be a baseline measurement taken for each metric at the same time for consistency. Baseline measurement refers to a non-smart device data collection for the purposes of this paper. Then, once all data collection had ended, it would be entered into a programmed emulation of a patient portal to enter collected biometrics and to eventually model the generated data from the smart health device's vs the baseline hand generated measurements. The program itself is a basic user interface that takes in each of the collected measurements for smart health and baseline measurements separately. The data points would then plot in a line graph for users to see and evaluate the consistency of performance for each smart device/metric over the period of two weeks. This would allow users to use this aid to analyze performance and make a final determination as to the reliability for a particular smart health device.

However, as stated above, due to unforeseen health circumstances… I was unable to conduct the outdoor portion of data collection and needed to switch to simulated data collection entirely. Therefore, I kept the planning details within this project report to be able to outline the empirical plan in place originally for the project but update and make conclusions on the basis of simulated data.

*A. Planned Exact Data Collection Distributions*

Data collection was planned to be conducted in such a manner so that the yielded results would present a balanced look into whether smart devices provided consistent and accurate (therefore reliable) performance in a wide range of network and environmental conditions over time.

- Indoor/2.4GHz: 7 Collections

- Outdoor/2.4GHz: 7 Collections
- Indoor/5GHz: 7 Collections
- Outdoor/5GHz: 7 Collections

IV. EVALUATION

The experiment that was planned to be conducted in this investigation was a biometric data collection over the course of two weeks under varying network and environmental conditions. The health metrics that were planned to be collected in this project include: blood pressure, temperature (Fahrenheit), and heart rate. The metrics were to be collected by a smart blood pressure monitor, smart thermometer, and smart watch respectively. In addition, there was a plan to have a baseline measurement taken at the same time as every other metric is taken, for each of the metrics. Collections were planned to be taken twice a day, every day, for two weeks. Therefore, there were 28 total smart device collections planned for each metric, and 28 corresponding baseline measurements planned for each metric as well. Data collection distribution plans were outlined in section III part A.

*A. Devices Planned to be Used and Set Up*

The smart devices planned to be used in this project included:

1) *Smart Thermometer*
2) *Smart Blood Pressure Device*
3) *Smart Watch (heart rate measurement only)*

Each device had to be researched to ensure that it performed the specified function and was a smart device that utilized an internet based connection. Due to the unforeseen health circumstances outlined above, they were not used for the data in the end for this project but are relevant for future work.

The baseline measurements for each metric were planned to be taken as follows, heart rate: hand count, blood pressure: non-internet based blood pressure monitor, temperature: non-internet based thermometer.

The planned use of varying network strengths for data collection meant that I needed to test and set up a dual band router to ensure that each network strength was functioning accurately and that each smart device could receive the appropriate network strength when needed. The network strengths planned to be tested in this paper were 5GHz and 2.4GHz through the use of the use of the dual band router to alternate strengths.

*B. Data Collection*

Data collection was planned to have taken place across the span of two weeks, twice a day for each of the outlined metrics: blood pressure, heart rate, and temperature. More specifically, data collection was planned to occur at each day once at 8AM and once at 8PM for consistency. Data collection was planned as a balanced distribution when it came to the environmental condition and network strength collected in as outlined in section III part A. Additionally, data collections were planned to be conducted using both the smart devices and the outlined baseline collection methods for each measurement type.

Balanced data collection allows for a fair analysis on the consistency, functionality, and accuracy of the smart health devices. Also, it removes potential bias or skewing of the results through the use of single option environment and/or network strength evaluation. Therefore, the data can be relevant for a much wider set of environmental and network conditions.

Baseline measurements are not generally impacted by environment or network strength. Nonetheless they were still planned to be collected at the same time as the smart health device data collections to ensure consistent measurements.

Instead, my study utilized simulated data for each of the outlined metrics to be able to simulate reliability analysis under this planned investigative outline.

*C. Data Entry & Patient Portal Program*

In the original program plan, python was the original intended language for the patient portal program. However, due to complications in having an extablished graphical user interface (outlined in detail below in issued encountered), the program instead is a console based Java application instead. The program welcomes the user and collects the 28 data samples for each of the metrics of blood pressure, temperature, and heart rate for both the smart health and baseline data collections, which in this project were all simulated. Total number of collections from the program would then be 28 * 3 * 2 = 168.

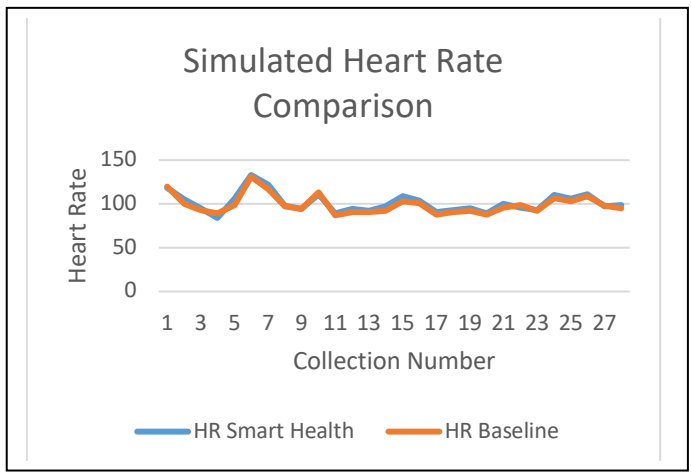

Fig. 1. Heart Rate comparison between the simulated smart health data and baseline data.

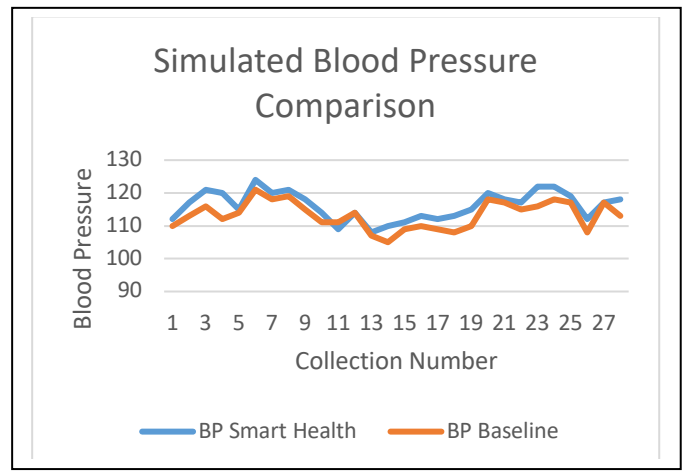

Fig. 2. Blood Pressure comparison between the simulated smart health data and baseline data.

The intention for the program was to be able to take the user data and visualize it in the form of graphs so that users can have a starting point to conduct analysis and determine if their smart health device is reliable in the varied conditions and when compared to the baseline measurements. However, due to additional complications in the Java packages required for statistical modelling and graphical representation, the program itself will not be able to have that functionality at this time (issue outlined in issues encountered section). To compensate and provide further visual aid, I have provided the charted results from the conducted simulations in excel generated charts instead. Intended future work for the program is for it to fully incorporate this feature instead of having to go to a third party to generate the charts. Then users would be able to review their collected data and make a determination regarding their smart device.

D. *Experiment Results, Raw Data Collected, and Associated Charts*

Full table of all simulated data located past references and notated by *.

Based on the simulated results from the comparison tables above, the first thing that stands out is that the heart rate comparison chart indicates that the simulated heart rate smart health device is consistent with the simulated baseline data. This indicates that the simulated smart health smartwatch could be considered a consistent tracker for heart rate. The simulated blood pressure comparison indicated that the simulated smart blood pressure device was the most inconsistent of the simulated smart health devices in terms of simulated daily measurement comparison.

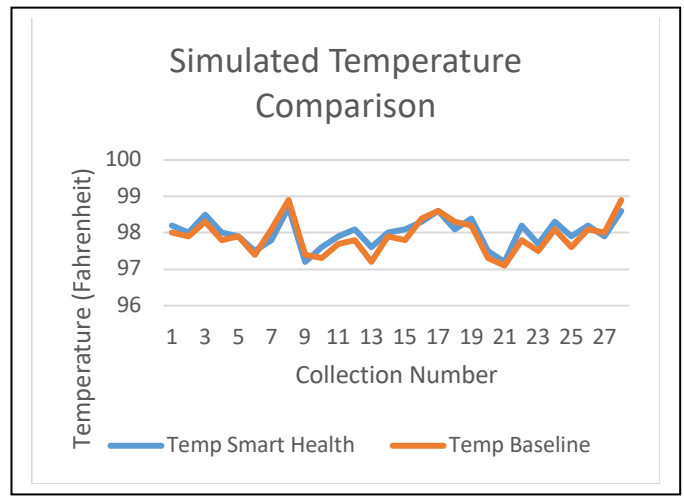

Fig. 3. Temperature comparison between the simulated smart health data and baseline data.

Therefore, based on the results, the simulated smart blood pressure device can be considered too inconsistent for use, and it might be a better option for a user to switch to a different machine (if it were a real device). The simulated temperature comparison also indicated that the simulated smart temperature device was consistent enough with the simulated baseline measurements. The simulated smart temperature device also demonstrates that it can be considered reliable (if it were a real device).

The fact that simulated data generated three different types of graphs/trend patterns demonstrates that the outlined/planned method of collection and analysis offers a solid starting point for in depth analysis on the consistency and reliability of the smart health devices.

## V. Issues Encountered

Originally, this project was planned as an investigative process with an empirical approach in experimental data collection. Unfortunately, due to health circumstances that were unforeseen, I had to switch to a simulated collection instead. I decided to keep the experiment data collection planning details to outline a fair and balanced planned approach that can offer a fair probe into smart health reliability. The simulated results offer only a sample of the potential of this approach and the impact on the ability to analyze smart health devices for reliability.

The original plan of this project had the programming portion of the project as a basic python interface with statistical modelling. However, due to issues in configuring the standard python graphics packages, such as tkinter, and the limitations of python when it comes to strict data management and data encapsulation, I made the decision to create the program in Java instead. The Java program offered a more basic terminal/console-based program that allowed for easy maintenance data management and configuration.

Finally, statistical modelling was made slightly difficult due to having to figure out how to populate multiple line charts in a single application window. The stored data structure was easy to work with itself, just translating that into a visualization in Java took was researched thoroughly. Therefore, due to additional complications in integrating the JavaFX and chart visualization add-ons, I was not able to integrate the visual element of the chart creations from the program. So, excel generated charts from the simulated data were included in this report instead to visualize what a future final result should look like for the program.

## VI. Conclusion

The expanding world of smart health is exciting and full of possibility for improving the access to high quality healthcare and user experience. This makes it more important than ever to be able to trust the smart health devices we use to perform consistently across changing and varied environments and network conditions. The accuracy of the data under these varied conditions is what makes a particular device reliable. Through this project I planned to empirically conduct an analysis on three smart health devices through varied network strengths and environmental conditions over the span of two weeks. Instead, I kept the plan as a framework for future work and switched to simulated data values instead. I have learned that the original plan of two weeks would not produce a large enough sample size to draw any outright conclusions on the performance of any single product. Also, that although my planned method of data collection was balanced and fair, that this method of collection was not optimal when it came to being able to have long periods of consistent collection in a single environment at a time vs constant variation. Constant variation is useful data, but it is more useful when paired with consistent collection, which allows an analyst to examine both cases separately and in conjunction. There were also clear limitations as to the amount of time there was to conduct this investigation, which traditionally should be done over the span of months to a year. Due to the time limitation and encountered health complication, I was not able to do physical data collections, and had to switch to simulated values instead due to health. The future work in this investigative frame would be to investigate larger sets of devices over more prolonged periods of time. Also, to be able to examine all features of a smart watch beyond just planning to examine heart rate. In the future, it would also be useful to improve the patient portal program to have a portable and robust GUI instead of the limited command line interaction that exists now. Program improvements could also incorporate the ability for multiple users to enter data for longer periods of time, better account security, or to just port the program to become a web based application for easy access. Additionally, the analysis can be improved to include more complex mathematical analysis that could feed into an AI program to get automated recommendations from the program as to whether the smart device in question is reliable or not so users can move on to another device if needed.

Overall, it was very fascinating to be able to formulate a project such as this one and it has a lot of potential for future use and improvement.